\documentclass[twocolumn,prl,superscriptaddress,showpacs]{revtex4}

\usepackage{amsmath,amssymb,graphicx,times}

\renewcommand\epsilon{\varepsilon}
\renewcommand\phi{\varphi}
\renewcommand\vec[1]{\boldsymbol{\mathrm{#1}}}
\newcommand\dotprod{\boldsymbol{\cdot}}
\newcommand\diff{\mathrm{d}}

\newcommand\expect[1]{\left\langle\vphantom{\big(}#1\right\rangle}

\newcommand\eq[1]{Eq.~\eqref{eq:#1}}
\newcommand\fig[1]{Fig.~\ref{fig:#1}}

\begin{document}
\addtolength{\textheight}{\baselineskip}

\title{Crossover in the Slow Decay of Dynamic Correlations in the Lorentz Model}
\author{Felix H{\"o}f\/ling}
\affiliation{Arnold Sommerfeld Center for Theoretical Physics (ASC)  and Center for
NanoScience (CeNS), Department of Physics,
Ludwig-Maximilians-Universit{\"a}t M{\"u}nchen, Theresienstra{\ss}e 37,
80333 M{\"u}nchen, Germany}
\affiliation{Hahn-Meitner-Institut,
Abteilung Theorie, Glienicker Stra{\ss}e 100, 14109 Berlin, Germany}
\author{Thomas Franosch}
\affiliation{Arnold
Sommerfeld Center for Theoretical Physics (ASC)  and Center for
NanoScience (CeNS), Department of Physics,
Ludwig-Maximilians-Universit{\"a}t M{\"u}nchen, Theresienstra{\ss}e 37,
80333 M{\"u}nchen, Germany}

\begin{abstract}
The long-time behavior of transport coefficients in a model for spatially heterogeneous media
in two and three dimensions is investigated by
Molecular Dynamics simulations. The behavior of the velocity
auto-correlation function is rationalized in terms of a
competition of the critical relaxation due to the underlying
percolation transition and the hydrodynamic power-law anomalies. In
two dimensions and in the absence of a diffusive mode, another
power law anomaly due to trapping is found with an exponent $-3$
instead of $-2$. Further, the logarithmic divergence of the Burnett
coefficient is corroborated in the dilute limit; at finite density, 
however, it is dominated by stronger divergences.
\end{abstract}


\maketitle

Spatial heterogeneities often give rise to intriguing slow dynamics in complex materials,
manifested for example by broad frequency-dependent relaxation processes in colloidal gels which form stress-sustaining networks close to the sol-gel transition~\cite{Dinsmore:2006,Ramos:2005,Sciortino:2004}.
Similarly, the presence of differently sized proteins, lipids and sugars in the cytoplasm of eukaryotes, summarized as cellular crowding, is identified by slow anomalous transport as its most distinctive fingerprint~\cite{Caspi:2002,Tolic-Norrelykke:2004,Weiss:2004}.
A further prominent example are sodium silicates, where the formation of a space-filling network of channels allows for slow diffusion of sodium ions in an arrested host matrix~\cite{Meyer:2004+Voigtmann:2006}.
A minimal model that encompasses spatial disorder and slow dynamics is provided by the Lorentz model~\cite{vanBeijeren:1982}, i.\,e.,
classical point particles explore without mutual interaction a
$d$-dimensional space in the presence of a frozen array of
randomly distributed (possibly overlapping) hard spherical obstacles of radius $\sigma$ and
concentration $n$.

Recently, striking behavior of the velocity auto-correlation function (VACF), $\psi(t) := v^{-2}
\expect{\vec{v}(t) \dotprod \vec{v}(0)}$, has been reported for a dense hard-sphere system $(d=3)$ close to the freezing transition~\cite{Williams:2006}. At intermediate time scales, a regime of anti-correlations emerges due to the well-known ``rattling'' of particles in their cages \cite{Rahman:1966+1964}. The long-time behavior exhibits an intriguing cross-over scenario from long-living positive correlations, $\psi(t)\simeq A_\text{fl} t^{-3/2}$, to a high-density regime characterized by slowly decaying anti-correlations, $\psi(t)\simeq -A_\text{fl}' t^{-5/2}$. The former corresponds to the celebrated long-time anomaly in simple liquids~\cite{Alder:1967+1970}, connected to the formation of a vortex pattern due to local momentum conservation. The mechanism for the latter decay is presumably of totally different origin: in an array of immobilized obstacles, the dynamics of a tagged particle always remembers its frozen cage---a mechanism well-known for the Lorentz model~\cite{vanBeijeren:1982}.



For the Lorentz model itself, there is a long-standing discrepancy between analytic theory and simulations about the manifestation of the long-time tail. The existence of the long-time anomaly in the Lorentz model has been predicted within a rigorous low-density expansion as $\psi(t) \simeq -A' t^{-d/2-1}$ for $t\to \infty$~\cite{Weijland:1968,Ernst:1971a}. Earlier computer simulations on two-dimensional systems~\cite{Bruin:1972+1974,Alder:1978+1983,Lowe:1993} identified a long-time relaxation of power-law type. At low densities, the expected exponent was confirmed; the amplitude, however, differed significantly from the theoretical prediction. At intermediate densities, the simulation results again suggest power-law behavior, which was described phenomenologically by non-universal, density-dependent exponents \cite{Alder:1978+1983,Alley:1979}.

The regime of higher obstacle densities poses considerable challenges for theory;
sequences of repeated ring collisions have been accounted for by a self-consistent variational repeated-ring theory
\cite{Masters:1982}.
\textcite{Goetze:1981+1982} have  developed a
mathematically consistent theory that covers the physics of the
low-density regime up to the predicted localization transition. In
particular, they predict a competition between a critical
power-law relaxation due to the fractal clusters at the percolation transition and the
universal long-time tail.
A similar scenario has been proposed for lattice variants of the Lorentz model \cite{vanVelzen:1988}, and there, evidence for a cross-over scenario has been reported \cite{Frenkel:1992}.

In this Letter, we present high-precision data  for the
two-dimensional overlapping Lorentz model for reduced obstacle densities, $n^*:=n\sigma^d$,
ranging from the dilute gas, $n^*= 0.005$, up to the percolation
threshold, $n^*_c \approx 0.35907$ \cite{Quintanilla:2000}, and deep into the localized phase.
In particular, we focus on the algebraic decay of the VACF at long times which is predicted for asymptotically low densities in dimensionless units as~\cite{Ernst:1971a}
\begin{equation}
\psi(s)  \simeq - \frac{n^*}{\pi} \frac{1}{s^2} \quad
\text{for}\quad s\to \infty, ~n^*\to 0, ~d=2,
\end{equation}
where $s=t/\tau$ is the mean number of collisions and $\tau^{-1} =
2n^* v/\sigma$ the collision rate~\footnote{Since the differential scattering cross section
, $\diff\sigma_{\text{sc}}(\theta)/\diff \theta = (\sigma/2) |\sin( \theta/2)|$,
is anisotropic for $d=2$, the collision rate differs from the momentum
relaxation rate, $1/\tau^* = 8n^* v/3\sigma$, that enters the
Boltzmann diffusion coefficient, $D_0 = v^2/2\tau^*$.},
and $v$ denotes the velocity of the particle.
In order to observe the universal tails, we analyze $\psi(t)$ up to times
corresponding to several $10^4$ collisions, which implies that a noise
level in the correlation of the order of $10^{-8}$(!) is required. Relying on an
event-oriented algorithm, we have simulated some $10^6$~trajectories per obstacle density, each trajectory covering
$10^6\!-\!10^8$~collisions.
We have calculated $\psi(t)$ by directly correlating velocities and obtain
accurate data up to a noise level of $10^{-5}$. Furthermore, we
have measured the mean-square displacement~(MSD), $\delta r^2(t) :=
\expect{\Delta \vec{R}(t)^2}$, and extracted the
diffusion coefficient $D$ from $\delta r^2(t\to\infty)\simeq 4Dt$.
 An alternative route to evaluate
$\psi(t)$ is to perform a numerical second derivative of $\delta
r^2(t)$. We have checked that both methods  yield identical results
within statistical errors. The second method further suppresses the noise
level by several orders of magnitude, up to a factor $10^3$. No
smoothening or sophisticated data manipulation has been used to obtain
the derivatives.

\begin{figure}
\includegraphics[width=\linewidth]{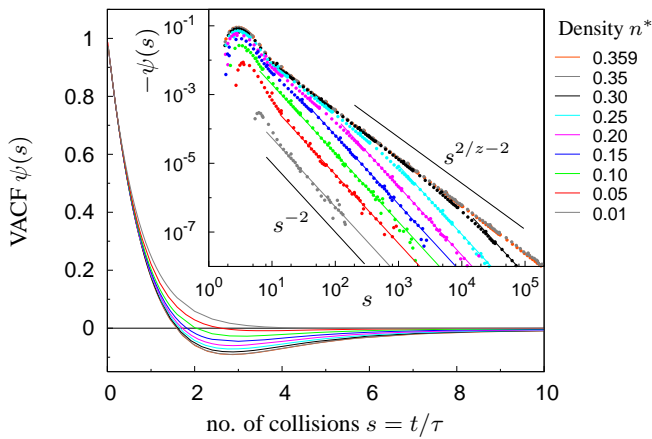}
\caption{(color online) Velocity autocorrelation function (VACF)
for the two-dimensional Lorentz model. The minimum is more pronounced
at higher densities. Inset: negative VACF on
double logarithmic scales. Solid lines are fits to the
universal long-time tails. The universal and the critical power
laws are indicated by thick straight lines, corresponding to exponents
$-2$ and $2/z-2\approx -1.34$.} \label{fig:vacf2d}
\end{figure}

Results for the VACF are shown in \fig{vacf2d} for
the full density range in the diffusive phase. On linear scales, the data for the lowest
density are indistinguishable from the exponential relaxation
$\exp(-4t/3\tau)$ of the Lorentz-Boltzmann theory. For intermediate
densities $(n^* \gtrsim 0.1)$, the VACF  enters the region of
anti-correlation  already after two collisions.
Since the diffusion coefficient is related to the total area under
the VACF by a Green-Kubo relation, $D = (v^2/d) \int_0^\infty
\psi(t) \diff t$, the areas of positive and negative region cancel
exactly at densities equal and above the percolation threshold,
$n^*_c$. In a double-logarithmic representation,
the data corroborate power-law behavior for
time windows covering 1$-$2 decades or, equivalently, 2$-$4 decades in
correlations. A gradual increase of the density towards $n^*_c$
gives rise to apparent density-dependent exponents, at least if
correlations below $10^{-4}$ are ignored. Careful inspection of
the VACF for $n^*=0.20$ reveals an intermediate power-law regime
as well as an universal long-time tail, consistent with the
competition of critical and universal relaxation predicted by
\textcite{Goetze:1981+1982}.

\begin{figure}
\includegraphics{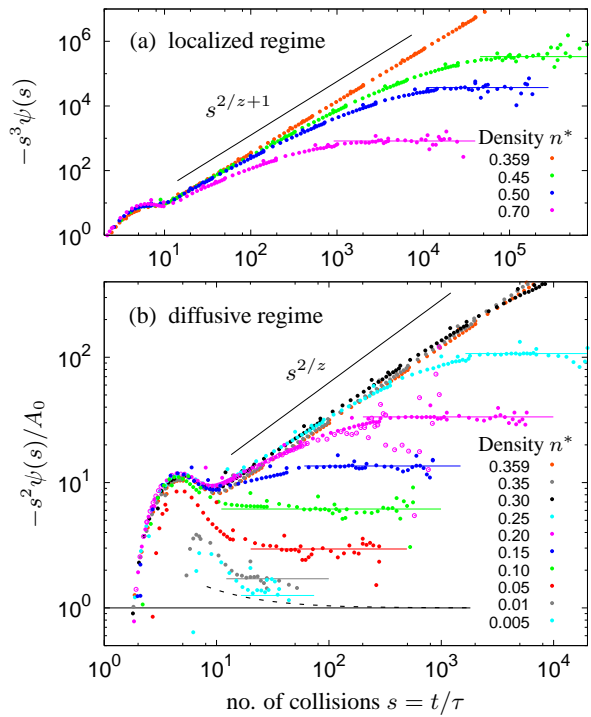}
\caption{(color online) Rectification of the universal
long-time tail (a) above and (b) below the percolation threshold.
Data points are obtained as numerical derivatives of the MSD; open circles for $n^*=0.20$ correspond to directly
correlating velocities; extracted amplitudes are indicated by solid lines.  The dashed line in (b) represents the leading correction to the tail, \eq{ltt_correction}.}
\label{fig:vacf-rescaled2d}
\end{figure}

As a most sensitive test for the crossover scenario,
\fig{vacf-rescaled2d} exhibits the VACF multiplied by the
expected power law $s^2$ of the universal tail. One infers that
$s^2 \psi(s)$ saturates at a constant in the accessible time
window for densities up to about~2/3 of the percolation threshold,
$n^* \lesssim 0.25$. For densities $n^*\lesssim 0.1$, the constant
is approached from above in qualitative agreement with the
prediction of \textcite{Das:1988} for the sub-leading long-time behavior,
\begin{equation}
\psi(s) \simeq - (n^*/\pi s^2)
 \left\{ 1 + 63/16 s + \ldots \right\}
\label{eq:ltt_correction}.
\end{equation}

 For higher densities, one observes an increase, following an apparent
density-dependent power law, before the universal tail is attained.
The time scale where the crossover occurs shifts to
longer times as the percolation threshold is approached, confirming the predicted scenario~\cite{Goetze:1981+1982}.
The density $n^*=0.10$ reaches its
asymptotic value remarkably early; this is due to a
cancellation effect of the sub-leading universal tail and
the onset of critical slowing down.

Very close to $n_c^*$, the VACF exhibits the critical relaxation
$\psi(t) \sim t^{2/z-2}$, which follows directly from the
prediction for the MSD, $\delta r^2(t) \sim
t^{2/z}$. The numerical value of the exponent,
$z\approx 3.03$~\footnote{A scaling relation, $z=(2\nu-\beta+\mu)/(\nu-\beta/2)$, connects $z$ to the geometric exponents $\nu=4/3$ and $\beta=5/36$ of lattice percolation in $d=2$ and the conductivity exponent $\mu=1.30$ describing the suppression of diffusion, $D\sim|n^*-n^*_c|^\mu$ \cite{Stauffer:Percolation}.}%
\newcounter{fnexponents}\setcounter{fnexponents}{\thefootnote},
coincides with results of simulations for diffusion on lattice percolation \cite{Stauffer:Percolation},
corroborating that the transport properties of the Lorentz model in
the close vicinity of the transition share the same universality
class~\footnote{To determine the exponent accurately,
one has to go to very long times, $t\gg 10^8\tau$, posing a
considerable challenge to computing. It appears that the critical
properties are masked by significant corrections to scaling
decaying even slower than in the three-dimensional case
\cite{Hoefling:2006,Hoefling:thesis}.}.

Long-time tails originating from power-law distributed exit rates
of the cul-de-sacs have been predicted even in the localized
regime~\cite{Machta:1985,Machta:1986a},  i.\,e., for obstacle
densities above $n_c^*$. In particular, the VACF should then decay as
$\psi(t) \sim t^{-3}$ for $n^*>n_c^*$ and $d=2$; a prediction
that has not been tested so far. Appropriate rectification plots
are included in \fig{vacf-rescaled2d}, and one infers that
the data follow such a power law for one order of magnitude in time,
i.\,e., three decades in correlation.

\begin{figure}
\includegraphics[width=\linewidth]{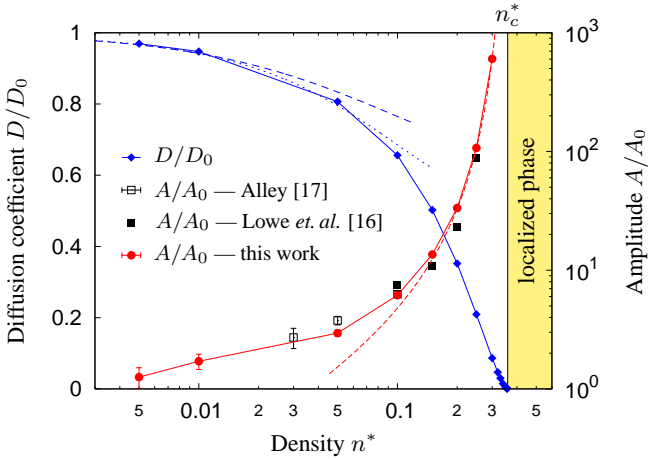}
\caption{(color online) Left axis (blue): suppression of the diffusion coefficient $D/D_0$ with respect to the Boltzmann-Lorentz
result $D_0$. The dashed line includes the leading low-density correction, the dotted line corresponds to \eq{diffusion}.
Right axis (red): reduced amplitude $A/A_0$ of the long-time tail from low densities up to the divergence at $n^*_c$. Dashed line: fit to \eq{A_D}, $A/A_0=1.6 n^*\, |n^*-n^*_c|^{-91/36}$.}
\label{fig:ltt-amplitudes2d}
\end{figure}

We have extracted dimensionless amplitudes $A$ of the
universal tail as the long-time limit of $-s^2 \psi(s)$. In the
regime of intermediate densities, our data are in semi-quantitative
agreement with earlier simulations \cite{Alley:1979,Lowe:1993},
see \fig{ltt-amplitudes2d}. As has been observed before, the values
of $A$ are  significantly larger than  the low-density prediction, $A_0
= n^*/\pi$. Since the calculation of $A_0$ relies on a perturbative
correction to the Boltzmann-Lorentz equation \cite{Ernst:1971a},
quantitative agreement requires the diffusion coefficient $D$ to be
sufficiently close to the Boltzmann value $D_0 = 3 v \sigma/16 n^*$.
As can be inferred from \fig{ltt-amplitudes2d}, this criterion is
not met even at low densities; a 40\% suppression of
$D/D_0$ occurs at $n^*=0.10$, signalling the onset of
subdiffusive motion. The diffusion coefficient for $n^*\leq 0.01$ is in
agreement with the first non-analytic correction  of
the low-density expansion \cite{Weijland:1968,Bruin:1972+1974}
\begin{equation}
\frac{D_0}{D} = 1 -\frac{4n^*}{3} \ln n^* - 0.8775\, n^* + 4.519\,
(n^* \ln n^*)^2.
\label{eq:diffusion}
\end{equation}
Although the amplitudes $A$ appear to approach the low-density prediction as $n^*\to 0$, the value for $A$ at $n^*=0.005$ still deviates by approximately 25\%.

Close to the percolation threshold, the crossover scenario suggests that the amplitude should actually diverge.
Matching the critical relaxation, $\psi(t)\sim t^{2/z-2}$, and the universal tail, $\psi(t)\simeq -A (t/\tau)^{-2}$,
at the divergent crossover time scale $t_*$ yields $\tau^2 A\sim t_*^{2/z}$. Assuming that $t_*$ also describes the crossover of the MSD from anomalous to diffusive transport, $t_*^{2/z}\sim D t_*$, entails the prediction,
\begin{equation}
\tau^2 A\sim D^{-2/(z-2)}\sim|n^*-n^*_c|^{-(2\nu-\beta)}
\label{eq:A_D}
\end{equation}
as $n^*\to n^*_c$, where $\beta$ and $\nu$ are percolation exponents~[\thefnexponents]. The rapid increase of the amplitudes follows this prediction remarkably well; even at $n^*=0.1$, \eq{A_D} deviates by less than 30\% from the simulation results, whereas the low-density prediction is off by a factor~6, see \fig{ltt-amplitudes2d}.

\begin{figure}
\includegraphics[width=\linewidth]{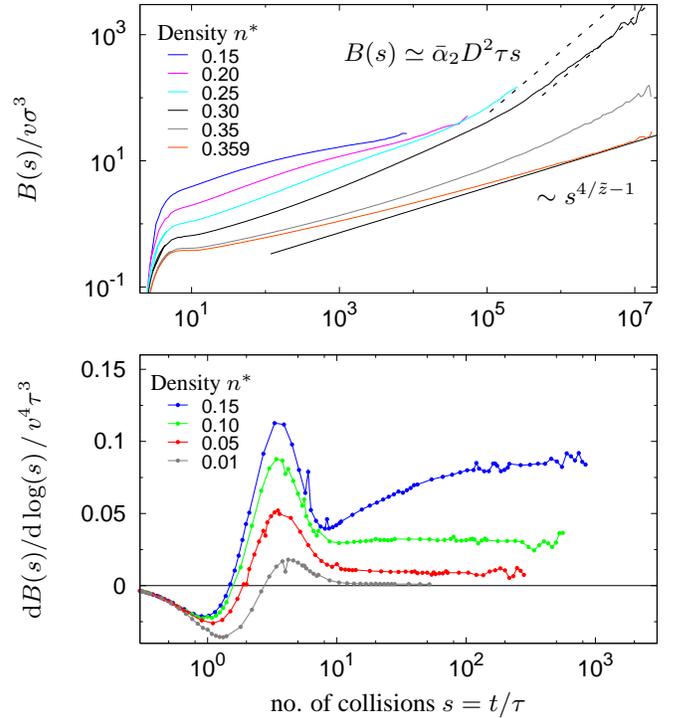}
\caption{(color online) Top: Critical (solid line) and linear (dashed
line) divergence of the Burnett coefficient at intermediate densities.
Bottom: The logarithmic divergence at low densities manifests itself as a finite long-time limit of $\diff B(s)/\diff \log(s)$.}
\label{fig:burnett2d}
\end{figure}

The VACF or, equivalently, the MSD is only the
simplest quantity exhibiting anomalous long-time behavior.
Deviations from Fickian diffusion are indicated by a non-vanishing
(super-)Burnett coefficient, which reads in two
dimensions
\begin{align}
B(t) &= \frac{1}{4!} \frac{\diff}{\diff t} \left[ \frac{1}{2}
\expect{\Delta \vec R(t)^4} - \expect{\Delta \vec R(t)^2}^2
\right].
\label{eq:burnett}
\end{align}
Within a hydrodynamic mode-coupling approach, it has been predicted
that the Burnett coefficient diverges logarithmically in $d=2$~\cite{Ernst:1984,Machta:1984}.
Indeed, $\diff B(t)/\diff \log(t)$ saturates at a constant for low densities, see \fig{burnett2d}.
Again close to $n_c^*$, one expects this behavior
to be masked by the critical relaxation. Dynamic scaling predicts a
power-law divergence of the mean-quartic displacement,
$\expect{\Delta \vec{R}(t)^4} \sim t^{4/\tilde{z}}$, where
$\tilde{z} = (2\nu-\beta + \mu)/(\nu-\beta/4)\approx2.95<z$~\cite{Hoefling:2006}.
 Then, the $B(t)$ increases according to $B(t)
\sim t^{4/\tilde{z}-1}$ at the critical density---consistent
with \fig{burnett2d}.
The presence of finite clusters renders the dynamics spatially heterogeneous, even below $n^*_c$.
A heterogeneous superposition of Gaussian processes yields a linearly divergent Burnett coefficient,
$B(t)\simeq \bar\alpha_2 D^2 t$ for $t\to\infty$, with $\bar \alpha_2=(4/3)(1/P_\infty -1)$, and $P_\infty$ denotes the fraction of mobile particles. We find semi-quantitative agreement with this prediction, see~\fig{burnett2d}. In the dilute limit, the prefactor is expected to vanish as $\bar\alpha_2 D^2\sim n^*$.



\begin{figure}
\includegraphics[width=\linewidth]{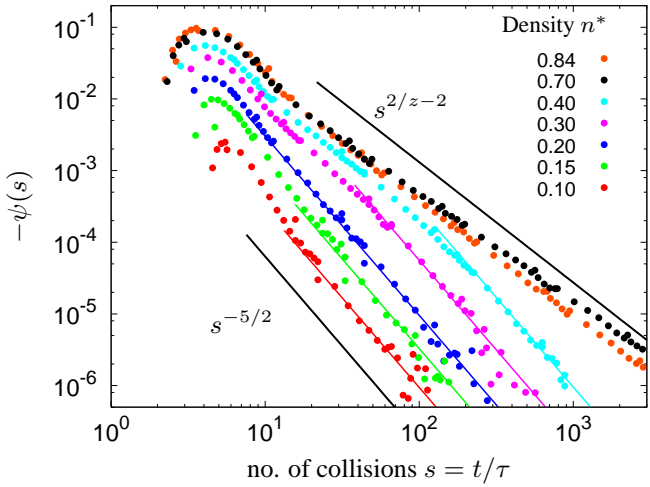}
\caption{(color online) VACF for the three-dimensional Lorentz model; a crossover similar to \fig{vacf2d} from the critical relaxation, $\psi(s)\sim s^{2/z-2}$, to the universal long-time tail, $\psi(s)\sim s^{-5/2}$, can be observed.}
\label{fig:vacf-ltt3d}
\end{figure}

Let us briefly comment on the the long-time behavior of the VACF
for the three-dimensional Lorentz model. Recently, the critical
properties of the localization transition have been analyzed in
terms of a scaling Ansatz for the van Hove correlation function
\cite{Hoefling:2006,Hoefling:thesis}. The exponent of the
universal tail is larger, and the amplitude depends even stronger
on the density~\cite{Ernst:1971a},
\begin{equation}
\psi(s) \sim - \frac{(3\pi)^{3/2}}{16}
(n^*)^2 s^{-5/2} \quad \text{for}\quad d=3
\end{equation}
where $s=t/\tau$ again, and the collision rate is given by
$\tau^{-1} = \pi n^* v/\sigma$ for $d=3$. Such a predicted
behavior is much more difficult to observe, and to the
best of our knowledge, only rudimentary evidence has been reported for
the existence of this phenomenon~\cite{Park:1989}. Our results (\fig{vacf-ltt3d}) suggest a similar crossover scenario as for $d=2$: the universal long-time tail with exponent $-5/2$ is preceded by a critical relaxation with exponent $2/z-2\approx -1.68$; the latter covers a growing time window upon approaching the localization transition.

Finally, we emphasize again the universality of the negative tail: it relies on the general mechanism that the particle can return in its frozen heterogeneous environment by reversing its path, thus remembering the presence of free volume~\cite{vanBeijeren:1982}.
For the hard-sphere fluid of Ref.~\cite{Williams:2006}, the role of the frozen environment is taken by long-lived cages in  the high-density regime.
Hence, the negative tail, $\psi(t)\simeq - A_\text{fl}' t^{-5/2}$, should emerge in an intermediate time window, $\tau \ll t \ll \tau_\alpha$, bounded by the time scales for momentum relaxation $\tau$ and structural relaxation $\tau_\alpha$; the slowing down of the latter is also reflected  in a rapid increase of the  viscosity of the fluid, $\eta\sim \tau_\alpha$.
At even larger times, the positive hydrodynamic tail,  $\psi(t) \simeq  A_\text{fl} t^{-3/2}$, is expected to follow although its amplitude  should vanish rapidly, $A_\text{fl} \sim \eta^{-3/2}$~\cite{Alder:1967+1970}.

We thank W.~G\"otze and E.~Frey for valuable discussions, and H.~van Beijeren for providing us with Ref.~\cite{Alley:1979}. F.\,H.\ acknowledges financial support from IBM Deutschland.


\begin{thebibliography}{45}
\expandafter\ifx\csname natexlab\endcsname\relax\def\natexlab#1{#1}\fi
\expandafter\ifx\csname bibnamefont\endcsname\relax
  \def\bibnamefont#1{#1}\fi
\expandafter\ifx\csname bibfnamefont\endcsname\relax
  \def\bibfnamefont#1{#1}\fi
\expandafter\ifx\csname citenamefont\endcsname\relax
  \def\citenamefont#1{#1}\fi
\providecommand{\bibinfo}[2]{#2}

\bibitem[{\citenamefont{Dinsmore \textit{et~al.}}(2006)\citenamefont{Dinsmore, Prasad,
  Wong, and Weitz}}]{Dinsmore:2006}
\bibinfo{author}{\bibfnamefont{A.~D.} \bibnamefont{Dinsmore}},
  \bibinfo{author}{\bibfnamefont{V.}~\bibnamefont{Prasad}},
  \bibinfo{author}{\bibfnamefont{I.~Y.} \bibnamefont{Wong}}, \bibnamefont{and}
  \bibinfo{author}{\bibfnamefont{D.~A.} \bibnamefont{Weitz}},
  \bibinfo{journal}{Phys. Rev. Lett.} \textbf{\bibinfo{volume}{96}},
  \bibinfo{pages}{185502} (\bibinfo{year}{2006}).

\bibitem[{\citenamefont{Ramos and Cipelletti}(2005)}]{Ramos:2005}
\bibinfo{author}{\bibfnamefont{L.}~\bibnamefont{Ramos}} \bibnamefont{and}
  \bibinfo{author}{\bibfnamefont{L.}~\bibnamefont{Cipelletti}},
  \bibinfo{journal}{Phys. Rev. Lett.} \textbf{\bibinfo{volume}{94}},
  \bibinfo{pages}{158301} (\bibinfo{year}{2005}).

\bibitem[{\citenamefont{Sciortino \textit{et~al.}}(2004)\citenamefont{Sciortino, Mossa,
  Zaccarelli, and Tartaglia}}]{Sciortino:2004}
\bibinfo{author}{\bibfnamefont{F.}~\bibnamefont{Sciortino}},
  \bibinfo{author}{\bibfnamefont{S.}~\bibnamefont{Mossa}},
  \bibinfo{author}{\bibfnamefont{E.}~\bibnamefont{Zaccarelli}},
  \bibnamefont{and}
  \bibinfo{author}{\bibfnamefont{P.}~\bibnamefont{Tartaglia}},
  \bibinfo{journal}{Phys. Rev. Lett.} \textbf{\bibinfo{volume}{93}},
  \bibinfo{pages}{055701} (\bibinfo{year}{2004}).

\bibitem[{\citenamefont{Caspi \textit{et~al.}}(2002)\citenamefont{Caspi, Granek, and
  Elbaum}}]{Caspi:2002}
\bibinfo{author}{\bibfnamefont{A.}~\bibnamefont{Caspi}},
  \bibinfo{author}{\bibfnamefont{R.}~\bibnamefont{Granek}}, \bibnamefont{and}
  \bibinfo{author}{\bibfnamefont{M.}~\bibnamefont{Elbaum}},
  \bibinfo{journal}{Phys. Rev. E} \textbf{\bibinfo{volume}{66}},
  \bibinfo{pages}{011916} (\bibinfo{year}{2002}).

\bibitem[{\citenamefont{Toli{\'c}-N{\o}rrelykke
  \textit{et~al.}}(2004)\citenamefont{Toli{\'c}-N{\o}rrelykke, Munteanu, Thon,
  Oddershede, and Berg-S{\o}rensen}}]{Tolic-Norrelykke:2004}
\bibinfo{author}{\bibfnamefont{I.~M.} \bibnamefont{Toli{\'c}-N{\o}rrelykke}},
  \bibinfo{author}{\bibfnamefont{E.-L.} \bibnamefont{Munteanu}},
  \bibinfo{author}{\bibfnamefont{G.}~\bibnamefont{Thon}},
  \bibinfo{author}{\bibfnamefont{L.}~\bibnamefont{Oddershede}},
  \bibnamefont{and}
  \bibinfo{author}{\bibfnamefont{K.}~\bibnamefont{Berg-S{\o}rensen}},
  \bibinfo{journal}{Phys. Rev. Lett.} \textbf{\bibinfo{volume}{93}},
  \bibinfo{pages}{078102} (\bibinfo{year}{2004}).

\bibitem[{\citenamefont{Weiss \textit{et~al.}}(2004)\citenamefont{Weiss, Elsner,
  Kartberg, and Nilsson}}]{Weiss:2004}
\bibinfo{author}{\bibfnamefont{M.}~\bibnamefont{Weiss}},
  \bibinfo{author}{\bibfnamefont{M.}~\bibnamefont{Elsner}},
  \bibinfo{author}{\bibfnamefont{F.}~\bibnamefont{Kartberg}}, \bibnamefont{and}
  \bibinfo{author}{\bibfnamefont{T.}~\bibnamefont{Nilsson}},
  \bibinfo{journal}{Biophys. J.} \textbf{\bibinfo{volume}{87}},
  \bibinfo{pages}{3518} (\bibinfo{year}{2004}).

\bibitem[{\citenamefont{Meyer \textit{et~al.}}(2004)\citenamefont{Meyer, Horbach, Kob,
  Kargl, and Schober}}]{Meyer:2004+Voigtmann:2006}
\bibinfo{author}{\bibfnamefont{A.}~\bibnamefont{Meyer}},
  \bibinfo{author}{\bibfnamefont{J.}~\bibnamefont{Horbach}},
  \bibinfo{author}{\bibfnamefont{W.}~\bibnamefont{Kob}},
  \bibinfo{author}{\bibfnamefont{F.}~\bibnamefont{Kargl}}, \bibnamefont{and}
  \bibinfo{author}{\bibfnamefont{H.}~\bibnamefont{Schober}},
  \bibinfo{journal}{Phys. Rev. Lett.} \textbf{\bibinfo{volume}{93}},
  \bibinfo{pages}{027801} (\bibinfo{year}{2004});
\bibinfo{author}{\bibfnamefont{T.}~\bibnamefont{Voigtmann}} \bibnamefont{and}
  \bibinfo{author}{\bibfnamefont{J.}~\bibnamefont{Horbach}},
  \bibinfo{journal}{Europhys. Lett.} \textbf{\bibinfo{volume}{74}},
  \bibinfo{pages}{459} (\bibinfo{year}{2006}).

\bibitem[{\citenamefont{van Beijeren}(1982)}]{vanBeijeren:1982}
for a review see
\bibinfo{author}{\bibfnamefont{H.}~\bibnamefont{van Beijeren}},
  \bibinfo{journal}{Rev. Mod. Phys.} \textbf{\bibinfo{volume}{54}},
  \bibinfo{pages}{195} (\bibinfo{year}{1982}).

\bibitem[{\citenamefont{Williams \textit{et~al.}}(2006)\citenamefont{Williams, Bryant,
  Snook, and van Megen}}]{Williams:2006}
\bibinfo{author}{\bibfnamefont{S.~R.} \bibnamefont{Williams}},
  \bibinfo{author}{\bibfnamefont{G.}~\bibnamefont{Bryant}},
  \bibinfo{author}{\bibfnamefont{I.~K.} \bibnamefont{Snook}}, \bibnamefont{and}
  \bibinfo{author}{\bibfnamefont{W.}~\bibnamefont{van Megen}},
  \bibinfo{journal}{Phys. Rev. Lett.} \textbf{\bibinfo{volume}{96}},
  \bibinfo{pages}{087801} (\bibinfo{year}{2006}).

\bibitem[{\citenamefont{Rahman}(1966)}]{Rahman:1966+1964}
\bibinfo{author}{\bibfnamefont{A.}~\bibnamefont{Rahman}}, \bibinfo{journal}{J.
  Chem. Phys.} \textbf{\bibinfo{volume}{45}}, \bibinfo{pages}{2585}
  (\bibinfo{year}{1966});
  \bibinfo{journal}{Phys. Rev.} \textbf{\bibinfo{volume}{136}},
  \bibinfo{pages}{A405} (\bibinfo{year}{1964}).

\bibitem[{\citenamefont{Alder and Wainwright}(1967)}]{Alder:1967+1970}
\bibinfo{author}{\bibfnamefont{B.~J.} \bibnamefont{Alder}} \bibnamefont{and}
  \bibinfo{author}{\bibfnamefont{T.~E.} \bibnamefont{Wainwright}},
  \bibinfo{journal}{Phys. Rev. Lett.} \textbf{\bibinfo{volume}{18}},
  \bibinfo{pages}{988} (\bibinfo{year}{1967});
  \bibinfo{journal}{Phys. Rev. A} \textbf{\bibinfo{volume}{1}},
  \bibinfo{pages}{18} (\bibinfo{year}{1970}).

%
%

\bibitem[{\citenamefont{Weijland and van Leeuwen}(1968)}]{Weijland:1968}
\bibinfo{author}{\bibfnamefont{A.}~\bibnamefont{Weijland}} \bibnamefont{and}
  \bibinfo{author}{\bibfnamefont{J.~M.~J.} \bibnamefont{van Leeuwen}},
  \bibinfo{journal}{Physica (Amsterdam)} \textbf{\bibinfo{volume}{38}},
  \bibinfo{pages}{35} (\bibinfo{year}{1968}).

\bibitem[{\citenamefont{Ernst and Weijland}(1971)}]{Ernst:1971a}
\bibinfo{author}{\bibfnamefont{M.~H.} \bibnamefont{Ernst}} \bibnamefont{and}
  \bibinfo{author}{\bibfnamefont{A.}~\bibnamefont{Weijland}},
  \bibinfo{journal}{Phys. Lett. A} \textbf{\bibinfo{volume}{34}},
  \bibinfo{pages}{39} (\bibinfo{year}{1971}).

\bibitem[{\citenamefont{Bruin}(1972)}]{Bruin:1972+1974}
\bibinfo{author}{\bibfnamefont{C.}~\bibnamefont{Bruin}},
  \bibinfo{journal}{Phys. Rev. Lett.} \textbf{\bibinfo{volume}{29}},
  \bibinfo{pages}{1670} (\bibinfo{year}{1972});
  \bibinfo{journal}{Physica} \textbf{\bibinfo{volume}{72}},
  \bibinfo{pages}{261} (\bibinfo{year}{1974}).

%
\bibitem[{\citenamefont{Alder and Alley}(1978)}]{Alder:1978+1983}
\bibinfo{author}{\bibfnamefont{B.~J.} \bibnamefont{Alder}} \bibnamefont{and}
  \bibinfo{author}{\bibfnamefont{W.~E.} \bibnamefont{Alley}},
  \bibinfo{journal}{J.~Stat. Phys.} \textbf{\bibinfo{volume}{19}},
  \bibinfo{pages}{341} (\bibinfo{year}{1978});
  \bibinfo{journal}{Physica A} \textbf{\bibinfo{volume}{121}},
  \bibinfo{pages}{523} (\bibinfo{year}{1983}).


\bibitem[{\citenamefont{Lowe and Masters}(1993)}]{Lowe:1993}
\bibinfo{author}{\bibfnamefont{C.~P.} \bibnamefont{Lowe}} \bibnamefont{and}
  \bibinfo{author}{\bibfnamefont{A.~J.} \bibnamefont{Masters}},
  \bibinfo{journal}{Physica A} \textbf{\bibinfo{volume}{195}},
  \bibinfo{pages}{149} (\bibinfo{year}{1993}).

\bibitem[{\citenamefont{Alley}(1979)}]{Alley:1979}
\bibinfo{author}{\bibfnamefont{W.~E.} \bibnamefont{Alley}}, Ph.\,D.\ thesis,
  \bibinfo{school}{California Univ.}, \bibinfo{address}{Davis}
  (\bibinfo{year}{1979}).

\bibitem[{\citenamefont{Masters and Keyes}(1982)}]{Masters:1982}
\bibinfo{author}{\bibfnamefont{A.}~\bibnamefont{Masters}} \bibnamefont{and}
  \bibinfo{author}{\bibfnamefont{T.}~\bibnamefont{Keyes}},
  \bibinfo{journal}{Phys. Rev. A} \textbf{\bibinfo{volume}{26}},
  \bibinfo{pages}{2129} (\bibinfo{year}{1982}).

\bibitem[{\citenamefont{G{\"o}tze \textit{et~al.}}(1981{\natexlab{a}})\citenamefont{G{\"o}tze, Leutheusser, and
  Yip}}]{Goetze:1981+1982}
\bibinfo{author}{\bibfnamefont{W.}~\bibnamefont{G{\"o}tze}},
  \bibinfo{author}{\bibfnamefont{E.}~\bibnamefont{Leutheusser}},
  \bibnamefont{and} \bibinfo{author}{\bibfnamefont{S.}~\bibnamefont{Yip}},
  \bibinfo{journal}{Phys. Rev. A} \textbf{\bibinfo{volume}{23}},
  \bibinfo{pages}{2634} (\bibinfo{year}{1981}{\natexlab{a}});
  \emph{ibid.} \textbf{\bibinfo{volume}{24}},
  \bibinfo{pages}{1008} (\bibinfo{year}{1981}{\natexlab{b}});
  \emph{ibid.} \textbf{\bibinfo{volume}{25}},
  \bibinfo{pages}{533} (\bibinfo{year}{1982}).

\bibitem[{\citenamefont{van Velzen \textit{et~al.}}(1988)\citenamefont{van Velzen,
  Ernst, and Dufty}}]{vanVelzen:1988}
\bibinfo{author}{\bibfnamefont{G.~A.} \bibnamefont{van Velzen}},
  \bibinfo{author}{\bibfnamefont{M.~H.} \bibnamefont{Ernst}}, \bibnamefont{and}
  \bibinfo{author}{\bibfnamefont{J.~W.} \bibnamefont{Dufty}},
  \bibinfo{journal}{Physica A} \textbf{\bibinfo{volume}{154}},
  \bibinfo{pages}{34} (\bibinfo{year}{1988}).

\bibitem[{\citenamefont{{Frenkel} \textit{et~al.}}(1992)\citenamefont{{Frenkel}, {van
  Luijn}, and {Binder}}}]{Frenkel:1992}
\bibinfo{author}{\bibfnamefont{D.}~\bibnamefont{{Frenkel}}},
  \bibinfo{author}{\bibfnamefont{F.}~\bibnamefont{{van Luijn}}},
  \bibnamefont{and} \bibinfo{author}{\bibfnamefont{P.-M.}
  \bibnamefont{{Binder}}}, \bibinfo{journal}{Europhys. Lett.}
  \textbf{\bibinfo{volume}{20}}, \bibinfo{pages}{7} (\bibinfo{year}{1992}).

\bibitem[{\citenamefont{Quintanilla \textit{et~al.}}(2000)\citenamefont{Quintanilla,
  Torquato, and Ziff}}]{Quintanilla:2000}
\bibinfo{author}{\bibfnamefont{J.}~\bibnamefont{Quintanilla}},
  \bibinfo{author}{\bibfnamefont{S.}~\bibnamefont{Torquato}}, \bibnamefont{and}
  \bibinfo{author}{\bibfnamefont{R.~M.} \bibnamefont{Ziff}},
  \bibinfo{journal}{J. Phys. A} \textbf{\bibinfo{volume}{33}},
  \bibinfo{pages}{L399} (\bibinfo{year}{2000}).

\bibitem[{\citenamefont{Das and Ernst}(1988)}]{Das:1988}
\bibinfo{author}{\bibfnamefont{S.~P.} \bibnamefont{Das}} \bibnamefont{and}
  \bibinfo{author}{\bibfnamefont{M.~H.} \bibnamefont{Ernst}},
  \bibinfo{journal}{Physica A} \textbf{\bibinfo{volume}{153}},
  \bibinfo{pages}{67} (\bibinfo{year}{1988}).

%
\bibitem[{\citenamefont{Stauffer and Aharony}(1994)}]{Stauffer:Percolation}
\bibinfo{author}{\bibfnamefont{D.}~\bibnamefont{Stauffer}} \bibnamefont{and}
  \bibinfo{author}{\bibfnamefont{A.}~\bibnamefont{Aharony}},
  \emph{\bibinfo{title}{Introduction to Percolation Theory}}
  (\bibinfo{publisher}{Taylor \& Francis}, \bibinfo{address}{London},
  \bibinfo{year}{1994}), \bibinfo{edition}{2nd} ed.

\bibitem[{\citenamefont{H{\"o}f\/ling \textit{et~al.}}(2006)\citenamefont{H{\"o}f\/ling,
  Franosch, and Frey}}]{Hoefling:2006}
\bibinfo{author}{\bibfnamefont{F.}~\bibnamefont{H{\"o}f\/ling}},
  \bibinfo{author}{\bibfnamefont{T.}~\bibnamefont{Franosch}}, \bibnamefont{and}
  \bibinfo{author}{\bibfnamefont{E.}~\bibnamefont{Frey}},
  \bibinfo{journal}{Phys. Rev. Lett.} \textbf{\bibinfo{volume}{96}},
  \bibinfo{pages}{165901} (\bibinfo{year}{2006}).

\bibitem[{\citenamefont{H{\"o}f\/ling}(2006)}]{Hoefling:thesis}
\bibinfo{author}{\bibfnamefont{F.}~\bibnamefont{H{\"o}f\/ling}}, Ph.\,D.\ thesis,
  \bibinfo{school}{Ludwig-Maximilians-Universit{\"a}t},
  \bibinfo{address}{M{\"u}nchen} (\bibinfo{year}{2006}).

\bibitem[{\citenamefont{Machta and Moore}(1985)}]{Machta:1985}
\bibinfo{author}{\bibfnamefont{J.}~\bibnamefont{Machta}} \bibnamefont{and}
  \bibinfo{author}{\bibfnamefont{S.~M.} \bibnamefont{Moore}},
  \bibinfo{journal}{Phys. Rev. A} \textbf{\bibinfo{volume}{32}},
  \bibinfo{pages}{3164} (\bibinfo{year}{1985}).

\bibitem[{\citenamefont{Machta}(1986)}]{Machta:1986a}
\bibinfo{author}{\bibfnamefont{J.}~\bibnamefont{Machta}}, \bibinfo{journal}{J.
  Stat. Phys.} \textbf{\bibinfo{volume}{42}}, \bibinfo{pages}{941}
  (\bibinfo{year}{1986}).

\bibitem[{\citenamefont{Ernst \textit{et~al.}}(1984)\citenamefont{Ernst, Machta,
  Dorfman, and van Beijeren}}]{Ernst:1984}
\bibinfo{author}{\bibfnamefont{M.~H.} \bibnamefont{Ernst}},
  \bibinfo{author}{\bibfnamefont{J.}~\bibnamefont{Machta}},
  \bibinfo{author}{\bibfnamefont{J.~R.} \bibnamefont{Dorfman}},
  \bibnamefont{and} \bibinfo{author}{\bibfnamefont{H.}~\bibnamefont{van
  Beijeren}}, \bibinfo{journal}{J. Stat. Phys.} \textbf{\bibinfo{volume}{34}},
  \bibinfo{pages}{477} (\bibinfo{year}{1984}).

\bibitem[{\citenamefont{Machta \textit{et~al.}}(1984)\citenamefont{Machta, Ernst, van
  Beijeren, and Dorfman}}]{Machta:1984}
\bibinfo{author}{\bibfnamefont{J.}~\bibnamefont{Machta}},
  \bibinfo{author}{\bibfnamefont{M.~H.} \bibnamefont{Ernst}},
  \bibinfo{author}{\bibfnamefont{H.}~\bibnamefont{van Beijeren}},
  \bibnamefont{and} \bibinfo{author}{\bibfnamefont{J.~R.}
  \bibnamefont{Dorfman}}, \bibinfo{journal}{J. Stat. Phys.}
  \textbf{\bibinfo{volume}{35}}, \bibinfo{pages}{413} (\bibinfo{year}{1984}).


\bibitem[{\citenamefont{Park and MacElroy}(1989)}]{Park:1989}
\bibinfo{author}{\bibfnamefont{I.-A.} \bibnamefont{Park}} \bibnamefont{and}
  \bibinfo{author}{\bibfnamefont{J.~M.~D.} \bibnamefont{MacElroy}},
  \bibinfo{journal}{Molecular Simulation} \textbf{\bibinfo{volume}{2}},
  \bibinfo{pages}{105} (\bibinfo{year}{1989}).
\end{thebibliography}
\end{document}